\documentclass[prl,twocolumn,showpacs,letter]{revtex4-1}

\usepackage{amsmath,amsfonts}
\usepackage{graphicx}  
\usepackage{wasysym}
\usepackage{amsmath,amssymb,psfrag,textcomp}
\usepackage{amsthm,upgreek,bm}   
\usepackage{bbm}

\newcommand{\bI}{\ensuremath{\mathbf{I}}}
\newcommand{\bG}{\ensuremath{\mathbf{G}}}

\newcommand{\bL}{\ensuremath{\mathbf{L}}}

\newcommand{\bB     }{\mbox{\boldmath$B$}}

\newcommand{\bzero  }{\mbox{\boldmath$0$}}

\newcommand{\bA     }{\mbox{\boldmath$A$}}

\newcommand{\bOne     }{\mbox{\boldmath$1$}}

\newcommand{\bpi     }{\mbox{\boldmath$\pi$}}

\newcommand{\blambda   }{\mbox{\boldmath$\lambda$}}

\usepackage{color}

\begin{document}
\title{Spectra of sparse non-Hermitian random matrices: an analytical solution}
\author{I. Neri$^{1,2}$, F. L. Metz$^{3}$} 
\affiliation{ 
${^1}$Universit\'e Montpellier 2, Laboratoire Charles Coulomb UMR 5221,
F-34095, Montpellier, France\\
${^2}$CNRS, Laboratoire Charles Coulomb UMR 5221, F-34095, Montpellier,
France \\
$^{3}$ Dipartimento di Fisica, Sapienza Universit\`{a} di Roma, INFN,
Sezione di Roma I, IPFC-CNR, P.le A. Moro 2, I-00185 Roma, Italy
}

\begin{abstract}
We present the exact analytical expression for the spectrum of a sparse
non-Hermitian random matrix ensemble, generalizing
two standard results in random-matrix theory: this analytical expression
constitutes
a non-Hermitian version of the Kesten-McKay measure as well as a sparse
realization
of Girko's elliptic law. 
Our exact result
opens  new perspectives in the study of several
physical problems modelled on sparse random graphs which are locally tree like.
In this context, we show
analytically that the convergence rate of a transport process on a very sparse
graph depends in a nonmonotonic way upon the degree of symmetry of the graph
edges.

\end{abstract} 
\pacs{02.50.-r, 02.10.Yn, 89.75.Hc}
\maketitle

\paragraph{Introduction}
Random matrices are an indispensable tool in the study of various
problems throughout physics and mathematics.  They pervade 
diverse areas as nuclear physics
\cite{Fyod}, number theory \cite{number}, quantum chaos and disordered
mesoscopic systems \cite{Guhr}, high-dimensional statistics
\cite{Johnstone} and information theory \cite{inf}, etc. Owing to the
ubiquitousness of random-matrix inspired models, exact analytical
results lead to further insights in several fields. 

The lions' share
of the research  on
random-matrix theory has been dedicated either to dense
ensembles or to sparse Hermitian ensembles (the
latter in the context of spectral graph theory). 
From the abundance of results known for dense ensembles \cite{Mehta, Bai, Bor}, 
we mention here the celebrated {\it Wigner's} law \cite{Wigner} and {\it Girko's
elliptic} law \cite{Girko} for, respectively, Hermitian and non-Hermitian
matrices. A large number of results for  spectra of graphs have also been
derived \cite{Mohar}. Probably, the most simple of these results 
is the {\it Kesten-McKay} measure for the spectra of undirected regular
graphs
\cite{Kesten}. Exact analytical expressions for spectra of graphs are important
 for processes on graphs: the computation
of effective resistances \cite{Klein},
synchronization in the presence of noise \cite{Wilk}, mixing times of transport
processes \cite{mixing},
bounds on learning in information theory \cite{Mont}, models for quantum chaos
\cite{chaos}, combinatorial problems on graphs \cite{Mck}, thermodynamics of
crystalline lattices \cite{Mcquarrie}, etc.

While many exact analytical results for sparse Hermitian
ensembles are known, exact expressions for the spectra of their non-Hermitian
counterparts have not yet been derived. 
This might be due to the difficulty of 
applying the standard manipulations from Hermitian
random matrix theory to the non-Hermitian case \cite{Bai, Bor}.
We refer to \cite{Fyod2, roger} as one of the few works which have
developed
exact results on sparse non-Hermitian matrices.  In \cite{Fyod2, roger}  Girko's
law has
been derived for sparse ensembles at high connectivities.  Despite these
efforts, analytical results are sparse and at
this moment even an equivalent of the paradigmatic Kesten-McKay law is not
known
for non-Hermitian ensembles. A non-Hermitian version of the Kesten-McKay measure
would be an important tool in the development of an exact description of
certain physical processes taking
place on 
graphs with oriented edges.  This is especially true if this measure would 
interpolate continuously between a fully
undirected and a fully directed graph.
As examples of areas which could
benefit from such a result, we  mention
biased diffusion processes \cite{Driven}, synchronization \cite{An}, neural
networks \cite{Leh,
Som1988} and non-Hermitian quantum systems, such as tight-binding models with
imaginary
vector potentials \cite{Hat, Ef} and quantum dissipative systems \cite{dis}.

In this work 
we present an exact analytical formula for the spectrum of a
sparse non-Hermitian random matrix ensemble, opening perspectives
in several fields which benefit from random matrix theory.
Such an exact result is possible
thanks to recent advances in the theory of sparse random matrices with a
local-tree like structure \cite{roger, Bordenave,
roger2}.   Our expression reduces to Girko's elliptic
law in its highly connected limit and to the Kesten-McKay law in its Hermitian
limit. 
Hence, it is a non-Hermitian equivalent of the Kesten-McKay law and a sparse
realization of Girko's elliptic law. 
Our result is remarkably
simple and allows to study
the evolution of the spectrum as a function of
the degree of symmetry in the graph edges. 
We illustrate the interest of this non-Hermitian Kesten-McKay law 
through one physical application: the exact calculation of the convergence
rate of a stochastic diffusion on a partially-oriented random graph. This
process can be
seen as a toy model for
vehicular traffic as well as biological transport \cite{Driven}.

\paragraph{The resolvent equations}
The complex eigenvalues
$\left\{\lambda_1, \lambda_2, \cdots,
\lambda_N\right\}$ of a non-Hermitian random matrix $\bA_N$ drawn from an
ensemble are defined as the
roots of the polynomial
$p(\lambda) =
{\rm
det}\left[\bA_N - \lambda \bI\right]$ of degree $N$ in $\lambda$.
We define the spectrum of the ensemble of random matrices $\bA_N$
at a certain point $\lambda = x + i y$ by
\begin{eqnarray}
\rho(\lambda) \equiv
\lim_{N\rightarrow \infty}N^{-1}\sum^N_{i=1}\delta_{\lambda_i\left(\bA_N
\right)},
\end{eqnarray}
assuming self-averaging of this quantity for $N\rightarrow \infty$.
The spectrum can be formally related to the resolvent
$\bG_{\bA}(\lambda)$ of $\bA$, defined by
$\bG_{\bA}(\lambda)
\equiv \left(\lambda - \bA\right)^{-1}$, through
the equation 
$\rho(\lambda) = \lim_{N\rightarrow \infty} (N \pi)^{-1}
\partial^{*} {\rm Tr} \bG_{\bA}(\lambda)$, where
$\partial^{*} = \frac{1}{2} \left(\frac{\partial}{\partial x} + i
\frac{\partial}{\partial y} \right)$ and
$(\dots)^{*}$ denotes complex conjugation. The
above equation follows from the identity
$\partial^{*} \lambda^{-1} = \pi \delta(x) \delta(y)$, which
can be proved by integrating both sides over
a small square centered at the origin. When
there is no ambiguity, we leave out the subindex $N$ in
matrices such as $\bA_N$.  

The resolvent $\bG_{\bA}(\lambda)$
is not properly defined at the eigenvalues
of $\bA$, which are distributed over the complex
plane. Therefore, one cannot apply the usual resolvent manipulations, as done for
Hermitian matrices \cite{Som1988}, to the non-Hermitian case. We can overcome
this problem through an Hermitization method
\cite{Girko90, Fein1997, Bor}. In this method one considers 
the resolvent $\bG_{\bB}(\eta)$
of the $2N \times 2N$ Hermitian block matrix $\bB_{2N}$
\begin{eqnarray}
 \bB_{2N} =   \left(\begin{array}{cc} \bzero_{N} & \bA_{N}-\lambda\\
\bA_{N}^\dagger-\lambda^\ast & \bzero_{N} \end{array}\right), \label{eq:block}
\end{eqnarray}
where $\bzero_{N}$ is an $N \times N$ matrix filled 
with zeros and $\eta \in \mathbb{R}$
is a regularizer which keeps $\bG_{\bB}(\eta)$ 
properly defined on the whole complex plane.
The resolvent $\bG_{\bA}(\lambda)$ follows
simply from the $N \times N$ lower-left block
of $\lim_{\eta \rightarrow 0} \bG_{\bB}(\eta)$, where
$\bG_{\bB}(\eta)$ are now computed using 
standard techniques for Hermitian 
matrices \cite{Bordenave}.

We  associate a graph to the matrix $\bA_N$ by connecting all sites $(i,j)$
with a
nonzero $A_{ij}$ element.  If this graph has a local tree structure, we have
the following exact expression for the spectrum 
$\rho(\lambda) = - \lim_{N\rightarrow \infty, \eta \rightarrow 0} (\pi N)^{-1}
\partial^{*} \sum_{i=1}^{N} [\bG_i]_{21}$,
where the $2 \times 2$ matrices $\{ \bG_i \}$
fulfill the closed set of equations,
\begin{eqnarray}
\bG^{-1}_i &=&-\blambda(\eta) - \sum_{\ell \in
\partial_{i}}\bA_{i\ell}\bG^{(i)}_\ell
\bA_{\ell i}, \label{eq:GiFirst} \\
(\bG^{(j)}_i)^{-1} &=& \bG^{-1}_{i} + \bA_{ij}\bG^{(i)}_j\bA_{ji} \,,
\label{eq:GiCFirst}
\end{eqnarray}
with $\blambda(\eta) = \left(\begin{array}{cc}i\eta&\lambda \\\lambda^\ast &
i\eta\end{array}\right)$ and $\bA_{i\ell} = \left(\begin{array}{cc}
0&A_{i\ell}
\\A^\ast_{\ell i}& 0
\end{array}\right)$. From the point
of view of random graphs, the matrix element $A_{\ell i}$ is 
the weight of the directed edge from node $\ell$ to $i$, while $\partial_{i}$
contains
the indices of the vertices belonging to the neighborhood
of node $i$. We could derive Eqs.~(\ref{eq:GiFirst}) and 
(\ref{eq:GiCFirst}) from using recursively the Schur-complement
formula \cite{Bordenave} and we refer the reader
to \cite{suppl1} for technical details.

Equations (\ref{eq:GiFirst}) and (\ref{eq:GiCFirst}) have been 
derived for the first time by Rogers and Castillo
\cite{roger} using an approach coming from the
theory of spin glasses. For most random matrix ensembles, these equations
have a highly intricate structure. To solve
them analytically is unfeasible without
further simplifications. 
The authors of \cite{roger} have
confirmed  exactness of Eqs.~(\ref{eq:GiFirst}-\ref{eq:GiCFirst}) by
finding an excellent agreement with 
direct diagonalization methods  for sparse
matrices with a local tree structure. 
In contrast, here we provide an
analytical solution to these equations for a particular
graph structure, which leads
to an explicit formula for $\rho(\lambda)$.

\begin{figure}[h!]
\includegraphics[ scale=0.6]{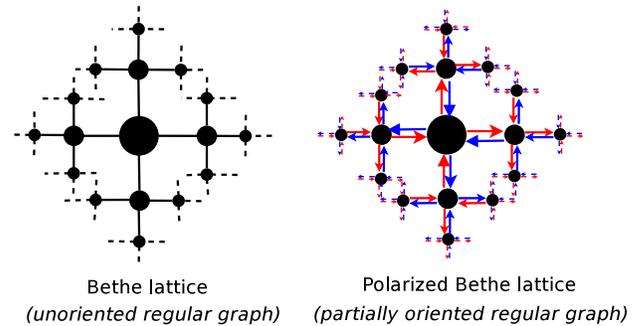}
\caption{(color online). An example of a Bethe lattice and a polarized Bethe
lattice, both with degree $k=2$. The matrix elements
corresponding with the red (blue) lines are given by $A_+ (A_-)$. 
}\label{fig:pol}
\end{figure}

In order to extract an interesting solution from 
Eqs.~(\ref{eq:GiFirst}) and (\ref{eq:GiCFirst}), capable 
to interpolate smoothly between a fully undirected and
a fully directed graph,
we now consider  an ensemble of random sparse non-Hermitian
matrices $\bA_N$ of dimension $N\times N$ having the following
constraints on its nondiagonal complex entries $A_{ij}$  
\begin{itemize}
 \item $(A_{ij},A_{ji}) \in \left\{(A_+,A_-), (A_-,A_+),
(0,0)\right\}$,
\item $\sum_{j(\neq i)}\delta(A_{ij}, A_{\pm}) = \sum_{j(\neq i)}\delta(A_{ji},
A_{\pm})= k$,
\end{itemize}
with $k>1$ and the diagonal elements $A_{ii}=0$.   This random ensemble can be
represented by a graph of vertices $[1..N]$ and directed segments
connecting adjacent vertices $(i,j)$ when $A_{ij}\neq 0$.
In the corresponding graph every pair of adjacent vertices is connected
by two edges in such a way that each vertex contains
$k$ incoming (outgoing) edges of weight $A_{+}$ ($A_{-}$)
and $k$ outgoing (incoming) edges of weight $A_{-}$ ($A_{+}$).  Hence, the
disorder in this ensemble is due to the topology of the underlying graph, while
the elements $A_{ij}$ are fixed to either of two values $A_{\pm}$. 
When the entries are real we have a polarized Bethe lattice, illustrated in
Fig.~\ref{fig:pol}. 
For $A_-=A_+$ we recover the unoriented Bethe lattice \cite{Baxter} of
degree $2k$, while 
for $A_-=0$ we obtain a fully oriented Bethe lattice
of in- and out-degree equal to $k$.  

The resolvent equations 
(\ref{eq:GiFirst}) and (\ref{eq:GiCFirst}) for this 
non-Hermitian ensemble become
\begin{eqnarray}
 \bG^{-1} &=& -\blambda(\eta) - k(\bA_- \bG_+
\bA_+ +\bA_+ \bG_- \bA_-), 
\label{eq:Gi} \\
 \bG^{-1}_{\pm} &=& \bG^{-1}+\bA_{\pm}\bG_{\mp}\bA_{\mp},
\label{eq:GiC}
\end{eqnarray}
where $\bA_\pm= \left(\begin{array}{cc}0 & A_{\pm} \\ A^\ast_{\mp} &
0\end{array}\right)$, and the spectrum 
follows from $\rho(\lambda) = - \pi^{-1} \lim_{\eta \rightarrow 0} \partial^{*}
\left[\bG\right]_{21}$.  The simplified equations Eqs. (\ref{eq:Gi}) 
and (\ref{eq:GiC}) follow from Eqs. (\ref{eq:GiFirst}) and
(\ref{eq:GiCFirst}) after considering the transitive  structure of
the polarized Bethe
ensembles for $N\rightarrow \infty$.  Equations (\ref{eq:Gi}) 
and (\ref{eq:GiC}) determine, for $N\rightarrow \infty$, the spectrum of 
a typical matrix drawn from our
ensemble, since the resolvent equations constitute  the typical local
neighborhoods of our ensemble. We now determine  the
solution to the equations (\ref{eq:Gi}) 
and (\ref{eq:GiC}).



\paragraph{The analytical expression for the spectrum}
In the following we use polar coordinates and set 
$A_{\pm}=p_{\pm} \exp(i\theta_{\pm})$. 
Solving Eqs. (\ref{eq:Gi}) and (\ref{eq:GiC}) we  
find the analytical expressions for $\bG$ and $\bG_{\pm}$,  see supplemental
material \cite{suppl2}. 
The spectrum $\rho_{0}(\lambda)$ for
real matrices $\bA_N$ (with $\theta_{+}=\theta_{-}=0$) is
given by
\begin{equation}
   \rho_{0} (\lambda) =
\frac{2kH p_{+}p_{-} \left[ \left(\frac{x}{S_{+}}
\right)^2 - \left(\frac
{y}{S_{-}}\right)^2\right] + C W   }{\pi
\left[\left(\frac{y}{S_-}
\right)^2+\left(\frac
{x}{S_+}\right)^2 + C \right]^2
Q_+Q_-} \label{eq:spectrum}
\end{equation}
for 
\begin{equation}
x^2 \: Q^{-2}_+ +
y^2 \: Q^{-2}_{-} < H^{-1},
\label{support}
\end{equation}
and $\rho_{0} (\lambda)=0$ otherwise. In Eqs.~(\ref{eq:spectrum}) and
(\ref{support}) we have
defined the constants $H$, $C$, $W$, $S_{\pm}$ and $Q_{\pm}$, which depend upon
$k$ and $p_{\pm}$ as follows:
\begin{eqnarray}
 2 H &=& k (p_{+}^2+p_{-}^2) +
\sqrt{k^2 (p_{+}^2 - p_{-}^2)^2 
  +4 (k-1)^2 \left( p_{+}  p_{-} \right)^{2} }, \nonumber \\
 C  &=& k^2 (k-1)^{-1} H^{-1} \left[   
(p_{+}^2+p_{-}^2)H - 2 \left( p_{+}  p_{-} \right)^{2} \right],  \nonumber \\
W&=&\left[H^2 + (2k-1)(p_{+}p_{-})^2 \right], \nonumber \\
Q_{\pm} &=& H \pm (2k-1)p_{+}p_{-}, \nonumber \\
  S_{\pm}^{2} &=& Q^{2}_{\pm} \left[
(H \mp p_{+}p_{-} )^2 - H C \right ]^{-1}.\nonumber 
\end{eqnarray}
Eqs.~(\ref{eq:spectrum}) and (\ref{support}) are the main result of our
work.
The support (\ref{support}) follows from a stability analysis of the trivial
solution to Eqs.~(\ref{eq:Gi}) and (\ref{eq:GiC}). Indeed,
for
large values of $\lambda$ we find
one stable trivial solution with $\rho_{0}(\lambda)=0$. This
solution 
is unstable at the support of $\rho_{0}(\lambda)$, see \cite{suppl2}. 
\begin{figure}[h!]
\begin{center}
\includegraphics[angle = -90, scale=0.3]{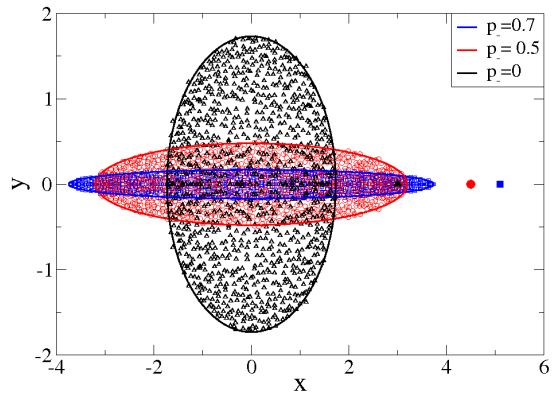}
\caption{(color online). Direct diagonalization results of matrices
of size
$\mathcal{O}(1e+3)$ (symbols) compared with the support 
(\ref{support}) (lines) for $k=3$, $p_{+} =1$ and
different values of $p_{-}$.}
\label{fig:cavitiesT}
\end{center}
\end{figure}  
In Figs.~\ref{fig:cavitiesT} and \ref{fig:cavitiesTT} we
compare direct diagonalization results
with, respectively, the analytical expressions for the
support (\ref{support}) and  the spectrum 
(\ref{eq:spectrum}). We find a very good correspondence
in both cases. Similar to Girko's law, the support forms 
an ellipse, but $\rho_{0}(\lambda)$ is 
non-uniform.

\begin{figure}[h!]
\begin{center}
\includegraphics[angle=-90, scale=0.23]{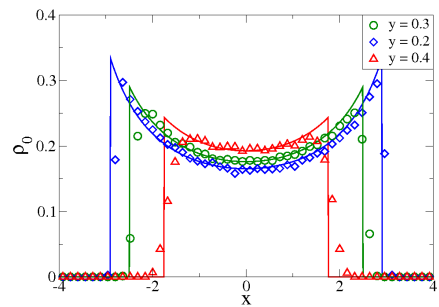}
\caption{(color online). Three cuts of $\rho_{0}(\lambda)$,
Eq.~(\ref{eq:spectrum}), along the real
direction are compared with 
direct diagonalization results of matrices of size
$\mathcal{O}(1e+3)$, averaged over $\mathcal{O}(1e+4)$ samples. The 
parameters are: $k=3$, $p_{+} = 1$, $p_{-}=0.5$.}
\label{fig:cavitiesTT}
\end{center}
\end{figure}

Remarkably, for
complex entries $A_{\pm}$ ($\theta_{\pm} \neq 0$) the spectrum $\rho(\lambda)$
and its support follow simply from Eqs.~(\ref{eq:spectrum}) and (\ref{support})
through a
clockwise rotation by an angle $\theta \equiv (\theta_{+} + \theta_{-})/2$ in
the
$(x,y)$-coordinate system, such that $\rho\left(\lambda \right) =  \rho_{0}
\left(\lambda 
e^{-i \theta} \right)$. Results for $\theta \neq 0$ are visualized in 
Fig.~\ref{fig:cavities}. 

\begin{figure}[h!]
\begin{center}
\includegraphics[scale=0.32]{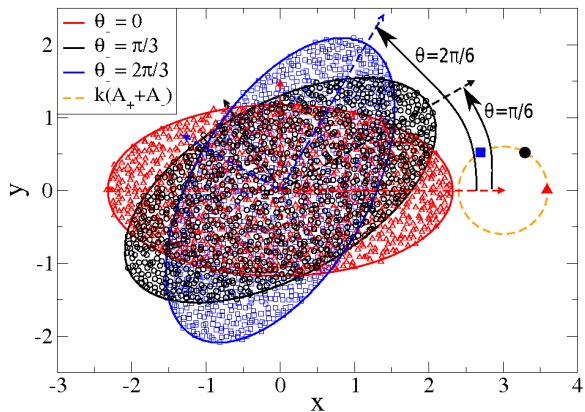}
\caption{(color online). Direct diagonalization results of
matrices of size $\mathcal{O}(1e+3)$ (symbols) compared with the 
rotated support of (\ref{support}) (solid lines) for $k=3$, 
$p_{+} = 1$, $\theta_{+} = 0$, $p_{-} = 0.2$
and different values of $\theta_{-}$. The
orange circle denotes the eigenvalues isolated from the bulk, given by
$k(A_{+} + A_{-})$. We notice the non-trivial rotation of the 
bulk spectrum of an angle $\theta$, different from the rotation of 
the isolated eigenvalue of an angle $2\theta$ around $k$.}
\label{fig:cavities}
\end{center}
\end{figure}

We discuss below a couple of interesting limiting cases:

{\it 1.$\,\,$Fully directed Bethe lattice}: This
limit is obtained by setting $p_{-} =0$ in
Eqs.~(\ref{eq:spectrum}) and (\ref{support})
\begin{eqnarray}
 \rho_{\rm DB}(\lambda) =
\frac{k-1}{\pi}\left(\frac{k \, p_{+}}{|\lambda|^2-k^2 p_{+}^2}\right)^2,
\label{eq:DB}
\end{eqnarray}
with the support $|\lambda|^2<k p_{+}^2$.
This formula has been conjectured before in
\cite{Metz2011, Bor}, but its rigorous proof remains an open problem
\cite{Bor}.

\vspace{0.2cm}
{\it 2.$\,\,$Undirected Bethe lattice}: By taking the limit
$p_{+} \rightarrow p_{-} \equiv p$ in Eqs.~(\ref{eq:spectrum}) and 
(\ref{support}), we obtain the Kesten-McKay law \cite{Kesten}
for a graph with degree $2k$
\begin{equation}
\rho_{\rm KM}(\lambda) = \delta(y)\frac{k
}{\pi}\frac{\sqrt{4 p^2 (2k-1)-x^2}}{4k^2 p^2-x^2},  
\end{equation}
with support $|x| < 2 |p| \sqrt{2k-1}$. For 
$\theta \neq 0$ and $p_{+} \rightarrow p_{-}$, the supports of the spectra
in Fig.~\ref{fig:cavities}
reduce to straight lines along the major axes of the ellipses, 
The projected spectrum is then given
by the Kesten-McKay measure. This concentration
of all eigenvalues on the straight line $y=x \tan(\theta)$ for
a given $\theta$ is
a general property of matrices of the type 
$\bA + \exp(i\theta) \bA^T$ ($[\bA]_{ij} \in \mathbb{R}$).
Therefore, the eigenvalues can
be brought to the real axis by a rotation.
Non-Hermitian matrices with a real spectrum have attracted
considerable attention as alternative theories 
to quantum mechanics \cite{Bender}.

\vspace{0.2cm}
{\it 3.$\,\,$Dense matrices}:
By rescaling  $p_{\pm} \rightarrow p_{\pm}/\sqrt{k}$ in Eqs.~(\ref{eq:spectrum})
and (\ref{support}), we 
obtain  the highly connected limit when $k\rightarrow \infty$
\begin{eqnarray} 
 \rho(\lambda) =
\frac{1}{\pi}\left(\frac{p^2_++p^2_-}{p^2_+-p^2_-}\right)^2,
\label{eq:FC}
\end{eqnarray}
with  support
 $\frac{x^2}{\left(p_++p_-\right)^2} + \frac{y^2}{\left(p_+-p_-\right)^2} <
\frac{1}{p^2_++p^2_-}$.
When we set $\mathbb{E}\left(A^2_{ij}\right) = 1$ and
$\mathbb{E}\left[A_{ij}A_{ji}\right] = \tau$, with  
$\mathbb{E}(\dots)$ denoting the ensemble average and $-1 \leq \tau \leq 1$,
Eq.~(\ref{eq:FC}) reduces to 
$ \rho(\lambda) =\left[ \pi(1-\tau^2) \right]^{-1}$ with
support $x^2/\left(1+\tau \right)^2 + y^2/\left(1-\tau\right)^2 < 2$
\cite{Som1988}. 
Thus, Eqs.~(\ref{eq:spectrum}) and (\ref{support}) form a non-trivial sparse 
realization of Girko's elliptic law.  Indeed, for $k\rightarrow \infty$ we have
highly connected sparse ensemble leading to Girko's elliptic law. Sparse
realizations of Wigner's semicircular law and Girko's circular law
has  recently been proven \cite{sparseWigner, GirkoCirc}.

\paragraph{Biased diffusion on  a regular graph}
As an application we determine the convergence rate to the
stationary state of a non-equilibrium
transport process on a regular graph. 
Consider a set of random
walkers moving along the edges of our partially-oriented regular graph model
with
$\theta_{\pm} = 0$ and transition rates $p_{\pm} > 0$.
The
relative occupancies $\bpi = (\pi_1, \pi_2, \cdots, \pi_N)$, with $\pi_i$ the
relative occupancy of the $i$-th site, fulfill the linear equation
\begin{eqnarray}
\frac{d}{dt}\bpi = \bL_{N} \bpi\:, \quad \bL_{N} = \bA_{N}-k(p_++p_-)\bOne_N \,.
\label{diffus}
\end{eqnarray}
Due to the Perron-Frobenius theorem, the Laplacian matrix $\bL_N$ has a
unique eigenvector with positive entries and
eigenvalue $\lambda_0 = 0$, which corresponds to the
stationary solution of Eq.~(\ref{diffus}).
All other eigenvalues have a negative
real part. If we order them as 
$\lambda_0>\mathcal{R}e\left(\lambda_1\right)>\cdots >
\mathcal{R}e\left(\lambda_{N-1}\right)$, then we define the spectral gap as
$g= |\mathcal{R}e\left(\lambda_1\right)|$. This diffusion process converges
exponentially to the steady state at a rate $g$. 

\begin{figure}
\begin{center}
\includegraphics[angle=-90, scale=0.25]{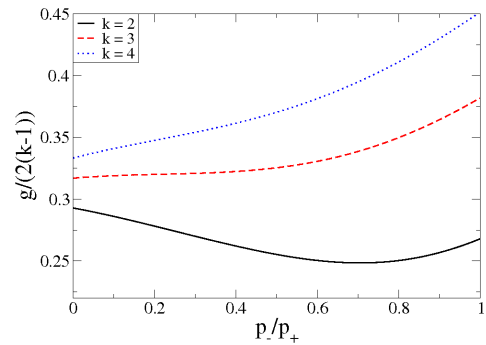}
\end{center}
\caption{(color online). Spectral gap as a function of the degree of symmetry
$\alpha = p_-/p_+$
for $p_+=1$ and
$k$ given. When $k\rightarrow \infty$, the
curves become straight lines.
}
\label{gap}
\end{figure}

 We use 
Eq.~(\ref{support}) for the boundary to obtain the spectral
gap for $N \rightarrow \infty$
\begin{eqnarray}
 \frac{g}{p_{+}} = (1+\alpha)k-\frac{\left[ h^2(\alpha, k)+(2k-1)\alpha
\right]}{h(\alpha, k) },
\end{eqnarray}
assuming that finite-size effects at the boundaries of $\rho_0 (\lambda)$
are negligible for large $N$ \cite{Bor}.
The quantity $\alpha = p_-/p_+$ is the degree of symmetry of the diffusion
process and
\begin{equation}
2h^{2}(\alpha, k) = k (1+ \alpha^2)
+\sqrt{k^2 (1 - \alpha^2)^2 + 4 (k-1)^2 \alpha^2 } \,.
\end{equation}
We plot $g$ as a function of $\alpha$ in Fig.~\ref{gap}.
Remarkably, the speed of convergence is non-monotonic as a function of 
$\alpha$ for $k=2$, indicating that for
intermediate values of $\alpha$ the diffusion process converges more
slowly to the steady state. For $k>2$ the fully symmetric process has the
fastest convergence. For dense random matrices such as the Ginibre ensemble the
spectral
gap is zero. Hence, we have determined analytically a physical
property absent in previous studies. 

\paragraph{Conclusions}
We have presented the analytical expression for the spectrum of a sparse
non-Hermitian random matrix ensemble, which reduces to the Kesten-McKay law for
Hermitian matrices and to the Girko's elliptic law in its highly connected
limit. 
Previous studies for sparse non-Hermitian random matrix ensembles relied on
numerical results.  Using a specific distribution for the matrix elements we
have found to our knowledge a first analytical expression for a non-Hermitian
random matrix ensemble generalizing the Kesten-Mckay measure.
Such a result 
 can stimulate
research in two different directions. On the one hand, we
have presented a random graph model which allows to address 
analytically how various processes on networks
depend on the orientation of the edges.
We have illustrated this through a study of the convergence
rate of a biased
transport process. Examples that deserve further study include 
network synchronization \cite{An}, neural networks \cite{Leh,
Som1988} and non-Hermitian quantum
mechanics  \cite{Hat, Ef, dis}. On the other hand, since 
our result is derived through a heuristic approach, it poses a
challenge for the development of new rigorous methods which deal with sparse
 non-Hermitian random matrices \cite{Bor}.  Further interesting research 
directions are: 
the development of exact results for  eigenvector localization \cite{Tar}
and  distribution of the largest eigenvalue \cite{Kab} of sparse
non-Hermitian
matrices.

\newpage
\clearpage
\setcounter{figure}{0}
\setcounter{table}{0}
\setcounter{equation}{0}
\setcounter{page}{1}

\setlength{\topmargin}{-0.5cm}
\setlength{\textheight}{22cm}
\setlength{\oddsidemargin}{0cm}
\setlength{\evensidemargin}{0cm}
\setlength{\textwidth}{15.8cm}
\setlength{\headsep}{0in}
\setlength{\parskip}{.15in}
\newcommand{\sectionsize}{\fontsize{12}{16}\selectfont}
\renewcommand{\bibfont}{\large}

\onecolumngrid

\begin{center}
{\huge\sf {Supplemental material 1}} \\
\vspace{0.4cm}
{\huge\sf {Derivation of the resolvent equations}} \\
{\LARGE\sf {I. Neri and F. L.  Metz}} \\
\end{center}

{\large

\section{\sectionsize Introduction}
Consider a large matrix $\bA_{N}$ of size $N$ with matrix elements
$A_{ij}=\left[\bA_N\right]_{ij} \in \mathbb{C}$.   In this document we present
the essential steps in deriving  the following
set of resolvent equations
\begin{eqnarray}
 \bG^{-1}_i &=&-\blambda(\eta) - \sum_{\ell \in
\partial
i}\bA_{i\ell}\bG^{(i)}_\ell
\bA_{\ell i}, \label{eq:GiFirstT} \\
(\bG^{(j)}_i)^{-1} &=& \bG^{-1}_{i} + \bA_{ij}\bG^{(i)}_j\bA_{ji},
\label{eq:GiCFirstT}
\end{eqnarray}
with $\blambda(\eta) = \left(\begin{array}{cc}i\eta&\lambda \\\lambda^\ast &
i\eta\end{array}\right)$ and $\bA_{i\ell} = \left(\begin{array}{cc}
0&A_{i\ell}
\\A^\ast_{\ell i}& 0
\end{array}\right)$. The set $\partial_i$ is defined as $\partial_i =
\left\{j\in[1..N]|A_{ji}\neq0 \vee A_{ij}\neq 0\right\}$. For more details 
and the definition of other quantities we refer to the main paper. 

The interest in the set of equations
(\ref{eq:GiFirstT}-\ref{eq:GiCFirstT}) is that it determines the spectrum of
$A_N$
through
\begin{eqnarray}
\rho(\lambda) = - \lim_{N\rightarrow \infty, \eta \rightarrow 0} \frac{1}{\pi N}
\partial^{*} \sum_{i=1}^{N} [\bG_i]_{21},
\end{eqnarray}
where $\partial^{*} = \frac{1}{2} \left(\frac{\partial}{\partial x} + i
\frac{\partial}{\partial y} \right)$.  It is also possible to determine other
quantities such as the diagonal correlator of eigenvectors \cite{Corrxx}, but we
do not elaborate on this point in this paper. 

The equations (\ref{eq:GiFirstT}) and
(\ref{eq:GiCFirstT}) have been derived for the first time in the recent work 
\cite{rogerxx}. However, a
rigorous derivation of these equations has not been presented yet.
The conjecture is that they are  
exact when $N\rightarrow \infty$ and when the underlying graph structure is
locally tree-like. By the
underlying graph we mean the connectivity graph one can form when connecting
all points $(i,j)$ which have $A_{ij}\neq 0$.  
The exactness of Eqs. (\ref{eq:GiFirstT}) and (\ref{eq:GiCFirstT}) 
is strongly supported by a very good agreement between their numerical 
solution and direct diagonalization methods \cite{rogerxx}.
The concept of a local-tree like
structure is properly defined in \cite{Aldousxx} and intuitively it means that
most
of the vertices have a finite neighborhood, where short loops are absent. This
is
true for randomly drawn graphs when they are very large \cite{Aldousxx}.  Of
course it is true that many real-world graphs are not constructed by a
random algorithm and contain an intricate loop structure. Nevertheless, it
remains interesting to consider random systems as a first model for 
complex systems.

In this supplemental material we present an alternative derivation of the set
of equations (\ref{eq:GiFirstT}-\ref{eq:GiCFirstT}). The main difference between
our method  
and the approach presented in \cite{rogerxx}
is that we avoid a mapping on a statistical mechanics problem. 
Therefore, the assumptions become more intuitive as we
avoid the factorization of a complex valued function \cite{rogerxx}, which
plays a role analogous to local marginals in spin models. 
Another advantage
of our approach is that it stands much closer to resolvent
methods used for rigorous spectral calculations on dense matrices \cite{Baixx}. 

\section{\sectionsize Hermitization procedure}
The spectral density of $A_N$ is obtained from
the resolvent $\bG_{\bA}(\lambda)
\equiv \left(\lambda - \bA\right)^{-1}$ through
 equation $\rho(\lambda) = \lim_{N\rightarrow \infty} (N \pi)^{-1}
\partial^{*} {\rm Tr} \bG_{\bA}(\lambda)$.
The first step of the method consists in mapping the calculation of the spectrum
of the non-Hermitian matrix
$\bA_N$ on a resolvent calculation of the Hermitian matrix $\bB_{2N}$
\cite{Fein1997xx}, defined as 
\begin{eqnarray}
 \bB_{2N} =   \left(\begin{array}{cc} \bzero_{N} & \bA_{N}-\lambda\\
\bA_{N}^\dagger-\lambda^\ast & \bzero_{N} \end{array}\right), \label{eq:blockT}
\end{eqnarray}
where $\bzero_{N}$ is a $N \times N$ matrix filled 
with zeros and $\eta \in \mathbb{R}$
is a regulator which keeps all quantities properly
defined.
In this framework, the spectrum follows from
\begin{eqnarray}
 \rho(\lambda) =- \left(N\pi\right)^{-1}\partial^{\ast}\lim_{\eta\rightarrow 0}
\rm{Tr}\left[\left(\begin{array}{cc} \bzero_{N}& \bOne_N\\\bzero_N
&\bzero_N\end{array} \right)\bG_{\bB}\left(\eta\right)\right ].
\label{spectrT}
\end{eqnarray}

The Hermitization procedure can also be presented graphically, as can
be seen in figure 1. 
\begin{figure}[h!]
\begin{center}
\includegraphics[ scale=0.2]{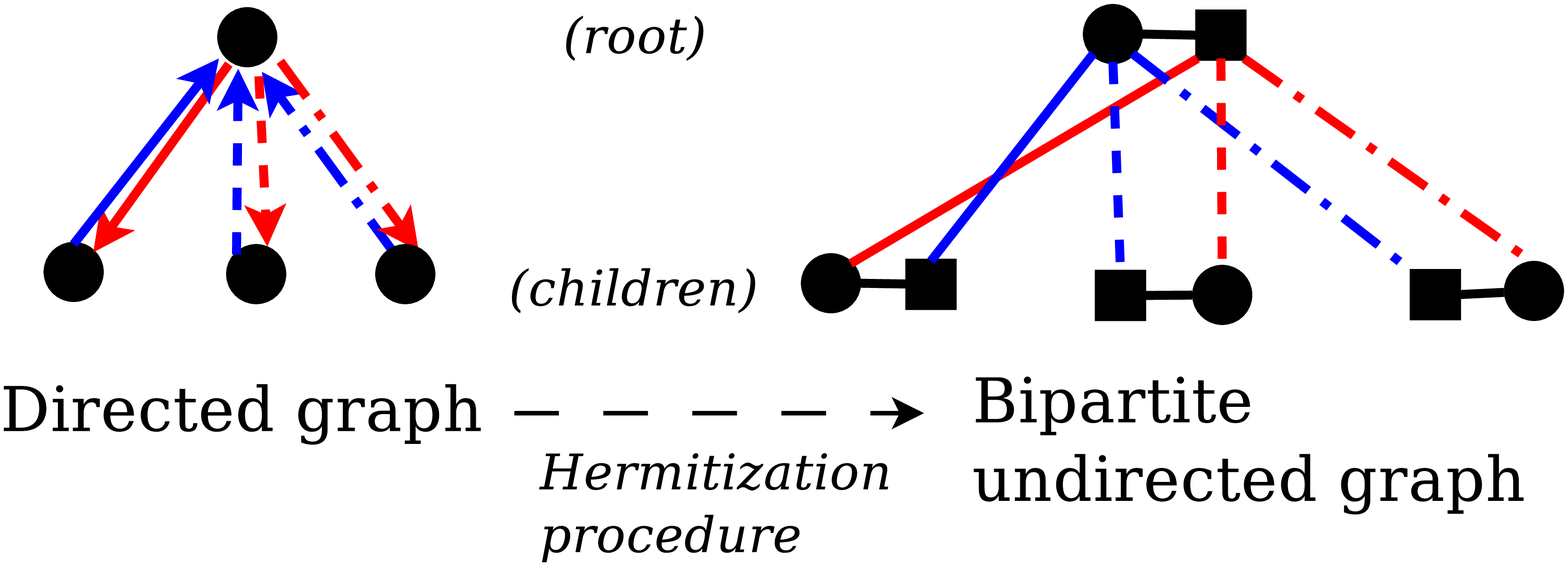}
\caption{(color online). Left: A non-Hermitian graph with oriented edges.
  Right: The undirected bipartite 
graph resulting from the application of the Hermitization
procedure \cite{Fein1997xx} to the subgraph on the left. } 
\label{fig:bipartite}
\end{center}
\end{figure}  
It corresponds to the translation of an oriented 
graph to an unoriented bipartite graph. Indeed, we can associate to the matrix
$\bA_N$ a graph $G = (V,E, W)$, with the set of vertices $V=[1,N]$ and the set
of
edges $E =
\left\{(i,j)|A_{ij}\neq 0\right\}$. Each edge has a weight
$w_{ij} = A_{ij}$, which is denoted by the mapping $W:E\rightarrow \mathbb{C}$. 
The edge is undirected when $w_{ij} = w^{\ast}_{ji}$, otherwise it is
directed. Therefore, the translation $\bA_{N}\rightarrow \bB_{2N}$
corresponds to a translation from an oriented graph to a bipartite unoriented
graph.

\section{\sectionsize Recursive application of the Schur-complement formula}
From equation (\ref{spectr}) we see that to calculate the spectrum, we need to
have an expression for the diagonal elements of the resolvent $\bG =
(\eta I_{2N}-\bB_{2N})^{-1}$.  Since this involves the inverse of a matrix, the
following
formula is very useful:
\begin{eqnarray}
  \left(\begin{array}{cc} A & B \\ C& D \end{array}\right)^{-1} &=&
\left(\begin{array}{cc} \left(A-BD^{-1}C\right)^{-1} &
-\left(A-BD^{-1}C\right)^{-1}BD^{-1} \\ -D^{-1}C\left(A-BD^{-1}C\right)^{-1}&
D^{-1} + D^{-1}C\left(A-BD^{-1}C\right)^{-1}BD^{-1} \end{array}\right).
\nonumber \\ \label{eq:Schur}
\end{eqnarray}
The above formula is referred to as the Schur-complement formula, and it is a
common
tool in the determination of spectra of dense matrices \cite{Baixx}.  Recently
the Schur-complement formula has also been applied to derive the
Kesten-Mckay law \cite{Bordenavexx}. Here we present a calculation similar
to the one presented in \cite{Bordenavexx}.  
  
\begin{figure}[h!]
\begin{center}
\includegraphics[ scale=0.5]{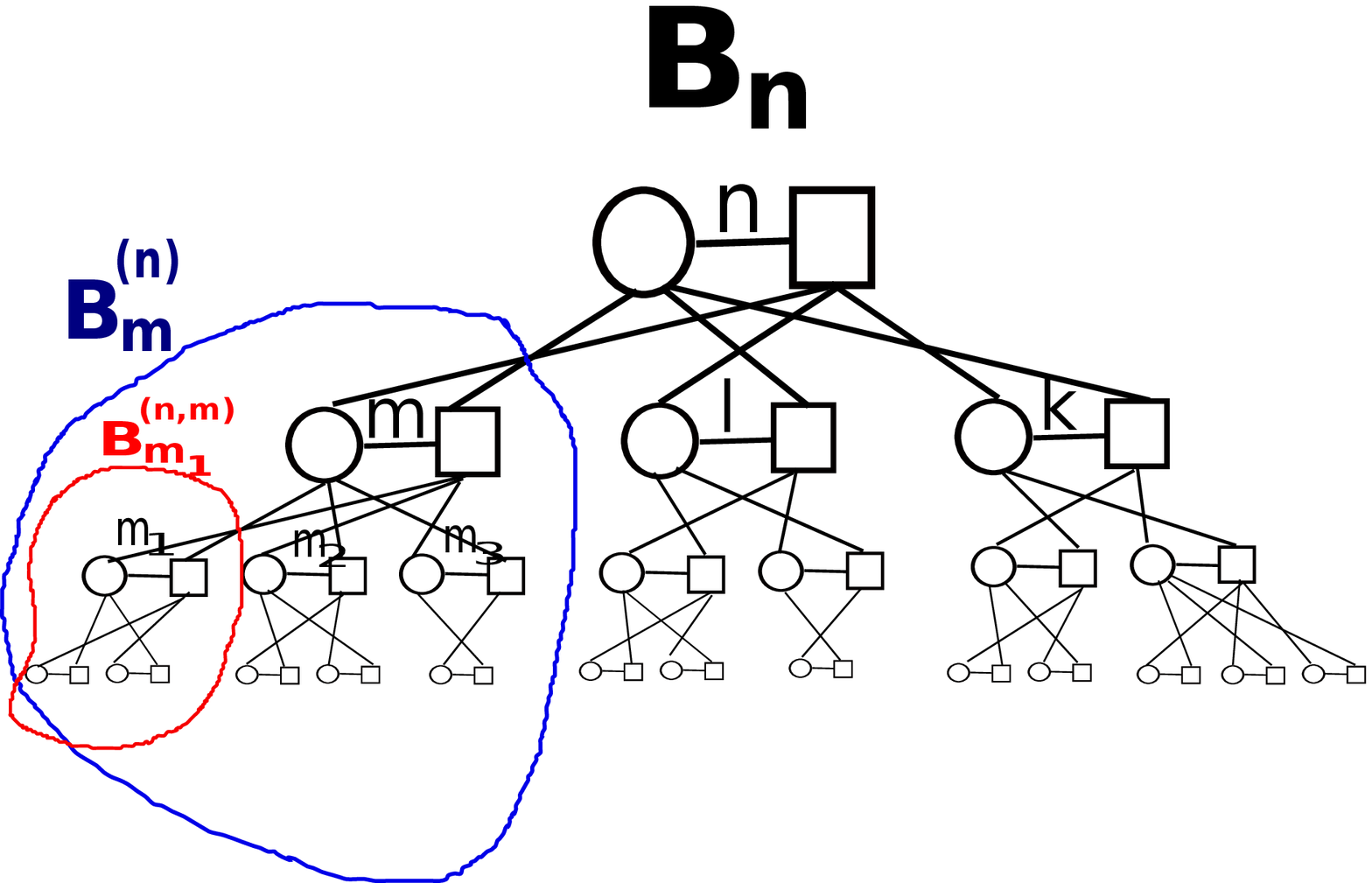}
\caption{A graphical representation of the recursive application of the
Schur-complement formula on the bipartite matrix $\bB_{2N}$.  We present here
the rooted graph around vertex $n$, corresponding to the matrix $\bB_{n}$.  We
can close the Schur-complement recursion by setting $\bB^{(m,n)}_{m_i} =
\bB^{(m)}_{m_i}$. }
\label{fig:cavities}
\end{center}
\end{figure}

When relabeling the indices of a matrix we do not change its
eigenvalues.  We use this property and relabel all the vertices such that
$(i,i+N)\rightarrow (2i, 2i+1)$.  It becomes useful to label each couple $(2i,
2i+1)$ by $i$ and write $\bB-\eta I$ as a matrix consisting of $2\times 2$
blocks. In
this matrix we have on the diagonal the $2\times 2$ matrix $\blambda(\eta) =
\left(\begin{array}{cc}i\eta&\lambda \\\lambda^\ast &
i\eta\end{array}\right)$ while the $ij$-th block is denoted by $\bA_{ij} =
\left(\begin{array}{cc}
0&A_{ij}
\\A^\ast_{j i}& 0
\end{array}\right)$.  We therefore find 
\begin{eqnarray}
 \bB'_{2N} = \left(\begin{array}{ccccc}-\blambda & \bA_{12} & \cdots
&\bA_{1(N-1)}& \bA_{1N} \\  \bA_{21} & -\blambda  & \cdots&\bA_{2(N-1)} &
\bA_{2N} \\
\\ \vdots & \vdots &   &\vdots & \vdots
\\  \bA_{(N-1)1} & \bA_{(N-1)2}  & \cdots &-\blambda& \bA_{(N-1)N}
\\
 \bA_{N1} & \bA_{N2} & \cdots & \bA_{N(N-1)}&-\blambda
\end{array}\right)
\end{eqnarray}
where the prime denotes the fact that we have relabeled the indices in
$\bB_{2N}$.  In fact, the matrix $\bB'_{2N}$ is the relabeled matrix following
from the above permutation operation on the matrix $\eta I-\bB_{2N}$.  
Let us consider now the unoriented graph $G' = (V',E',
W')$, with
$V'=[1..N]$, $E' = \left\{(i,j)|A_{ij}\neq 0\right\}$ and $w_{ij} = 1$ for all
edges.  We perform now a first depth search around a certain root vertex $n$ in
this unoriented graph and label the vertices accordingly. 
We define the rooted matrix $\bB^{(n)} = (\bB', n)$, as the matrix
which we have created from $\bB'$ through a permutation according to the
first-depth search around $n$.  We have therefore for $\bB^{(n)}$:
\begin{equation}
 \bB^{(n)}=\begin{pmatrix}
 -\blambda&\begin{matrix}\bA_{nn_1}&\bzero_2&\cdots\end{matrix}
&\begin{matrix}\bA_{nn_2}&\bzero_2&\cdots\end{matrix}
&\cdots&\begin{matrix}\bA_{nn_{k_n}}&\bzero_2&\cdots\end{matrix}\\
\begin{matrix}\bA_{n_1n}\\\bzero_2\\\vdots\end{matrix}&\bB^{(n)}_{n_1}
&&& \\
\begin{matrix}\bA_{n_2n}\\\bzero_2\\\vdots\end{matrix}&&\bB^{(n)}_{n_2}
&& \\
\vdots&&&&\\
\begin{matrix}\bA_{n_{k_n}n}\\\bzero_2\\\vdots\end{matrix}&&&&\bB^{(n)}_{k_n}
\end{pmatrix}.
\end{equation}
We have introduced the degree $k_n$, which denotes the number of matrix
elements $A_{ni}$, $i(\neq n)\in V$,  which  are different than zero.  The
matrices $\bB^{(n)}_{n_1}$ correspond to a subgraph of $G'$ associated to one
of the children nodes (or branches) of $n$.  This picture is visualized in
figure 1. 

We apply now the Schur-complement formula, to find:
\begin{eqnarray}
\left(\begin{array}{cc}\left[\left(\bB^{(n)}\right)^{-1}\right]_{11}&\left[
\left(\bB^{(n)}\right)^{-1}\right]_{12}\\
\left[\left(\bB^{(n)}\right)^{-1}\right]_{21}&\left[
\left(\bB^{(n)}\right)^{-1}\right]_{22}\end{array}\right) &=&
\frac{1}{-\blambda -
\sum^{k_n}_{j=1}\left( \bA_{nn_j} \bzero_2 \cdots \bzero_2 \right)
\left(\bB^{(n)}_{n_j}\right)^{-1}\left( \bA_{nn_j} \bzero_2 \cdots \bzero_2
\right)^\dagger} \nonumber \\ \label{eq:8}
\end{eqnarray}

We see that to determine the right hand side of equation (\ref{eq:8}), it is
necessary to have an expression in the upper $2\times 2$ dimensional block
of the inverse of the matrix $\bB^{(n)}_{n_1}$, i.~e., we need to know the
elements  $\left[\left(\bB^{(n)}_{n_1}\right)^{-1}\right]_{11}$,
$\left[\left(\bB^{(n)}_{n_1}\right)^{-1}\right]_{12}$,
$\left[\left(\bB^{(n)}_{n_1}\right)^{-1}\right]_{21}$ and
$\left[\left(\bB^{(n)}_{n_1}\right)^{-1}\right]_{22}$.  These elements can be
determined using again the Schur-complement formula (hence the recursion). 
Since $\bB^{(n)}_{n_1}$ is the matrix corresponding to one of the subgraphs
of $G'$, we have (setting $n_1=m$)
\begin{equation}
 \bB^{(n)}_{m}=\begin{pmatrix}
 -\blambda&\begin{matrix}\bA_{mm_1}&\bzero_2&\cdots\end{matrix}
&\begin{matrix}\bA_{jj_2}&0&\cdots\end{matrix}
&\cdots&\begin{matrix}\bA_{mm_{k_m}}&0&\cdots\end{matrix}\\
\begin{matrix}\bA^\dagger_{mm_1}\\0\\\vdots\end{matrix}&\bB^{(n,m)}_{m_1}
&&& \\
\begin{matrix}\bA^\dagger_{mm_2}\\0\\\vdots\end{matrix}&&\bB^{(n,m)}_{m_2}
&& \\
\vdots&&&&\\
\begin{matrix}\bA^\dagger_{mm_{k_m-1}}\\0\\\vdots\end{matrix}&&&&\bB^{(n,m)}_{
k_m-1}
\end{pmatrix}
\end{equation}
After applying again the Schur-complement formula, we can close the resultant
set of equations using 
$\bB^{(o)}_{m} = \bB^{(n,o)}_{m}$, for all $n\in \partial_o$ and $m\in
\partial_o$, with $\partial_o$.  Using this approximation we find the closed
set of equations (\ref{eq:GiFirstT}-\ref{eq:GiCFirstT}).    Indeed, setting 
\begin{eqnarray}
 \bG_n =
\left(\begin{array}{cc}\left[\left(\bB^{(n)}_m\right)^{-1}\right]_{11}&\left[
\left(\bB^{(n)}_m\right)^{-1}\right]_{12}\\
\left[\left(\bB^{(n)}_m\right)^{-1}\right]_{21}&\left[
\left(\bB^{(n)}_m\right)^{-1}\right]_{22}\end{array}\right)
\end{eqnarray}
and 
\begin{eqnarray}
 \bG^{(n)}_m =
\left(\begin{array}{cc}\left[\left(\bB^{(n)}_m\right)^{-1}\right]_{11}&\left[
\left(\bB^{(n)}_m\right)^{-1}\right]_{12}\\
\left[\left(\bB^{(n)}_m\right)^{-1}\right]_{21}&\left[
\left(\bB^{(n)}_m\right)^{-1}\right]_{22}\end{array}\right)
\end{eqnarray}
we recover indeed the equations (\ref{eq:GiFirstT}-\ref{eq:GiCFirstT}).

We remark that the
condition $\bB^{(o)}_{m} = \bB^{(n,o)}_{m}$ is indeed exact on a tree. 
Therefore, our approximation has some clear intuition.

}

\newpage
\clearpage
\setcounter{figure}{0}
\setcounter{table}{0}
\setcounter{equation}{0}
\setcounter{page}{1}

\setlength{\topmargin}{-0.5cm}
\setlength{\textheight}{22cm}
\setlength{\oddsidemargin}{0cm}
\setlength{\evensidemargin}{0cm}
\setlength{\textwidth}{15.8cm}
\setlength{\headsep}{0in}
\setlength{\parskip}{.15in}
\renewcommand{\bibfont}{\large}

\onecolumngrid

\begin{center}
{\huge\sf {Supplemental material 2}} \\
\vspace{0.4cm}
{\huge\sf {Solving the resolvent equations}} \\
{\LARGE\sf {I. Neri and F. L.  Metz}} \\
\end{center}

{\large

\section{\sectionsize Introduction}

In this supplemental material we present how the polarized Bethe lattice leads
to a non-trivial solution to the resolvent equations.  We present the exact
analytical solution to the resolvent equations in this case.  

\section{\sectionsize Resolvent equations: general case}
Let us consider a sparse non-Hermitian matrix $\bA$, with
$A_{ij}=\left[\bA\right]_{ij}$ and $i,j=1 \dots N$.  We define the set
$\partial_i=\left\{j|A_{ij}\neq 0\vee A_{ji}\neq 0 \right\}$.  
The resolvent equations in the $2\times 2$ matrices
$\bG_i$ and $\bG^{(i)}_{j}$
are given by
\begin{eqnarray}
 \bG^{-1}_i &=&-\blambda(\eta) - \sum_{\ell \in
\partial
i}\bA_{i\ell}\bG^{(i)}_\ell
\bA_{\ell i}, \label{eq:GiFirstTT} \\
(\bG^{(j)}_i)^{-1} &=& \bG^{-1}_{i} + \bA_{ij}\bG^{(i)}_j\bA_{ji}
\label{eq:GiCFirstTT}
\end{eqnarray}
with $\blambda(\eta) = \left(\begin{array}{cc}i\eta&\lambda \\\lambda^\ast &
i\eta\end{array}\right)$ and $\bA_{i\ell} = \left(\begin{array}{cc}
0&A_{i\ell}
\\A^\ast_{\ell i}& 0
\end{array}\right)$.  These equations present an algorithm which allows us to
determine, among other spectral quantities, the spectrum of the graph through
\begin{eqnarray}
 \rho(\lambda) = - \lim_{N \rightarrow \infty, \eta \rightarrow 0} 
\frac{1}{N \pi} \partial^{*} \sum^N_{i=1}[\bG_i]_{21}
\end{eqnarray}
where $\partial^{*} = \frac{1}{2} \left(\frac{\partial}{\partial x} + i
\frac{\partial}{\partial y} \right)$.  The equations (\ref{eq:GiFirstTT}) and
(\ref{eq:GiCFirstTT}) are conjectured to be exact when $N\rightarrow \infty$ and
when the graph is locally tree like.  Its use as an efficient and accurate
algorithm has been demonstrated in \cite{rogerx}.  
Here 
we present an analytical solution to
Eqs.~(\ref{eq:GiFirstTT}-\ref{eq:GiCFirstTT}) 
for a partially-oriented regular
graph.  Our formula generalizes Girko's elliptic law to the sparse case as
well as the Kesten-Mckay law to the non-Hermitian case. 


\section{\sectionsize Resolvent equations for partially-oriented regular graphs}
The resolvent equations for the partially-oriented regular graph, see figure 2
in the main paper, are given by

\begin{eqnarray}
 \bG^{-1} &=& -\blambda(\eta) - k(\bA_- \bG_+
\bA_+ +\bA_+ \bG_- \bA_-), 
\label{eq:GiT} \\
 \bG^{-1}_{\pm} &=& \bG^{-1}+\bA_{\pm}\bG_{\mp}\bA_{\mp}.
\label{eq:GiCT}.
\end{eqnarray}
These are equations (6-7) in the main paper. 
The set of equations (\ref{eq:GiT}-\ref{eq:GiCT}) has only one stable solution. 
This has been proven for the Hermitian case \cite{Bordenavex}, and we
conjecture,
for now, this to be also true in the non-Hermitian case.  Something which is
supported by numerically solving the resolvent equations.  We find for
large
values of $|\lambda|$ a trivial solution $(\bG^{\rm t}_{\pm}, \bG^{\rm t})$
(with $\rho(\lambda)=0$).   At small values of $|\lambda|$, the trivial solution
becomes unstable in favor of a non-trivial solution $(\bG^{\rm
n}_{\pm}, \bG^{\rm n})$ (with
$\rho(\lambda)>0$).   Below we describe the trivial solution $(\bG^{\rm
t}_{\pm}, \bG^{\rm t})$ and the non-trivial solution $(\bG^{\rm
n}_{\pm}, \bG^{\rm n})$.  We also determine the line in the complex plane
where the trivial solution becomes unstable.  This line corresponds with the
boundary of the support of the spectrum.  

Let us first define some constants which we will use throughout:
\begin{eqnarray}
U &\equiv& \frac{A_+A^{*}_- + A^{*}_+A_-}{2},  \nonumber 
\\ 
 V &\equiv&    \frac{A^{*}_+ A_+
+ A^{*}_- A_-}{2},  \nonumber  \\ 
W &\equiv&\frac{A_-A^{*}_- -A_+A^{*}_+}{2i},
 \nonumber  \\  
Z &\equiv& \frac{A_+A_- - A^{*}_+A^{*}_- }{2i}. \nonumber 
\end{eqnarray}
We remind that $A_+$ and $A_-$ are the non-zero matrix elements or equivalently
$(A_\pm, A_\mp)$ are the weights of the edges in the graph.  Every vertex is
incident to $k$ edges of weight  $(A_+, A_-)$ and $k$
edges of weight $(A_-, A_+)$. 
It is convenient to use polar coordinates and write down $A_+ =
p_+\exp(i\theta_+)$ and $A_-
= p_-\exp(i\theta_-)$.  

At last we point out that it will be useful to parametrize the resolvent
matrices as follow: 
\begin{eqnarray}
 \bG_{\pm}&=&  \left(\begin{array}{cc} a \pm b & c\\
 d& a\mp b\end{array}\right), \label{eq:1}\\ 
\bG &=& \left(\begin{array}{cc}  a'+b'&c' \\
d'& a'-b'\end{array}\right). \label{eq:2} 
\end{eqnarray}
with $(a,b,c,d)\in \mathbb{C}^4$.

\section{\sectionsize The trivial solution: $\rho(\lambda)=0$}
The trivial solution follows from substitution of
equations (\ref{eq:1}-\ref{eq:2}) in (\ref{eq:GiT}-\ref{eq:GiCT}) and setting
$a= b=0$.   The trivial solution is therefore given by
\begin{eqnarray}
 \bG^{\rm t}_\pm&=&  \left(\begin{array}{cc} 0 & c_0   \\
d_0&0\end{array}\right), \label{eq:3}\\ 
 \bG^{\rm t} &=& \left(\begin{array}{cc}  0&c'_0    \\
d'_0& 0\end{array}\right), \label{eq:4}
\end{eqnarray}
with
\begin{eqnarray}
c_0 &=&
\frac{-\lambda^{*}+\sqrt{\left(\lambda^{*}\right)
^2-4\left(2k-1\right)\left(U-iZ\right) } } {
2\left(2k-1\right)\left(U-iZ\right) }, \label{eq:c0} \\ 
d_0&=& \frac{-\lambda+\sqrt{\lambda^2 -
4\left(2k-1\right)\left(U+iZ\right)}}{2\left(2k-1\right)\left(U+iZ\right)}, 
\label{eq:d0}
\end{eqnarray}
and
\begin{eqnarray}
c'_0&=&  -\frac{2k-1}{\left(k-1\right)\lambda^{*} +
\sqrt{\left(\lambda^{*}\right)
^2-4\left(2k-1\right)\left(U-iZ\right) }},\nonumber \\ 
d'_0 &=&  -\frac{2k-1}{\left(k-1\right)\lambda +
\sqrt{\lambda
^2-4\left(2k-1\right)\left(U-iZ\right) }}. \nonumber
\end{eqnarray}
The spectrum follows from  $\rho(\lambda) = -\pi^{-1}\frac{\partial}{\partial
\lambda^{*}}d'_0 = 0$.

\section{\sectionsize The non-trivial solution: $\rho(\lambda)>0$}
The non-trivial solution follows from substitution of
equations (\ref{eq:1}-\ref{eq:2}) in (\ref{eq:GiT}-\ref{eq:GiCT}) and setting
$a\neq 0$ and $b\neq 0$:
\begin{eqnarray}
 a^2 &=& \left(c\:d
-\frac{1}{H} \right)
\left(\frac{\left(H-V\right)^2}{W^2 + \left(H-V\right)^2}\right), \nonumber \\
 b^2 &=&
-\left(c\:d
-\frac{1}{H} \right)
\left(\frac{W^2}{W^2 + \left(H-V\right)^2}\right), \nonumber  \\
 c&=& \frac{(2k-1)U \: \lambda^{*}   +
(Z(2k-1)-H)\lambda  }{H^2-(2k-1)^2(U^2+Z^2)},\nonumber \\
 d &=& \frac{- H \lambda^{*}+
\left(-iZ(2k-1)+(2k-1)U\right)\lambda
}{H^2-(2k-1)^2(U^2+Z^2)}, \nonumber
\end{eqnarray}
and for the resolvent:
\begin{eqnarray}
 (a')^2 &=&\tilde{M}^{-2}4k^2\Bigg[-\left(-x \left(-(2k-1)U + H\right) +
Z(2k-1)y \right)^2 
\nonumber \\ 
&& 
 -\left(y \left((2k-1)U + H\right) -
Z(2k-1)x\right)^2 
+\frac{H^2-(2k-1)^2\left(U^2+Z^2\right)}{H} \Bigg] \left[\frac{\left(V(H-V)
-W^2\right)^2}{W^2 +
\left(H-V\right)^2}\right], \label{eq:a}\nonumber \\
\nonumber \\ 
b' &=&0, \nonumber\\ 
c' &=&\tilde{M}^{-1}\Big[y
\left(2kZH\right) + x
\left(-H^2-(2k-1)\left(U^2+Z^2\right) 
+2kUH \right) \nonumber \\
&&
-i\left(y\left(H^2+(2k-1)\left(U^2+Z^2\right)+2kUH\right) +
x\left(-2kHZ\right)\right)\Big],\nonumber \\
d' &=& \tilde{M}^{-1}\Big[y
\left(2kZH\right) + x
\left(-H^2-(2k-1)\left(U^2+Z^2\right) 
+2kUH \right)
\nonumber \\
&&
+i\left(y\left(H^2+(2k-1)\left(U^2+Z^2\right)+2kUH\right) +
x\left(-2kHZ\right)\right)\Big]. \nonumber 
\end{eqnarray}
The denominator is given by  
\begin{eqnarray}
\tilde{M} = \frac{M}{H^2-(2k-1)^2\left(U^2+Z^2\right)},
\end{eqnarray}
with
\begin{eqnarray}
  M  &=&  4k^2\Bigg[-\left(-x \left(-(2k-1)U + H\right) +
Z(2k-1)y \right)^2 
\nonumber \\
&&  -\left(y \left((2k-1)U + H\right) -
Z(2k-1)x\right)^2 
+\frac{H^2-(2k-1)^2\left(U^2+Z^2\right)}{H} \Bigg] \left[\frac{\left(V(H-V)
-W^2\right)^2}{W^2 +
\left(H-V\right)^2}\right]
\nonumber \\
&& + \left[y \left(2kZH\right) + x
\left(-H^2-(2k-1)(U^2+Z^2) 
+2kUH \right) \right]^2 
\nonumber \\
 && +
\left[y\left(H^2+(2k-1)(U^2+Z^2)+2kUH\right) +
x \left(-2kHZ\right)\right]^2.
\end{eqnarray}
We remind that $\lambda=x+iy$. The spectrum of the non-trivial solution follows
then from  $\rho(\lambda) = -\pi^{-1}\frac{\partial}{\partial
\lambda^{*}}d'$
\begin{eqnarray}
 \rho\left(\lambda \right) =  \rho_{0}
\left(\lambda 
e^{-i \theta} \right) ,
\end{eqnarray}
with $\theta = \theta_++\theta_-$ and 
\begin{equation}
   \rho_{0} (\lambda) =
\frac{2kH p_{+}p_{-} \left[ \left(\frac{x}{S_{+}}
\right)^2 - \left(\frac
{y}{S_{-}}\right)^2\right] + C W   }{\pi
\left[\left(\frac{y}{S_-}
\right)^2+\left(\frac
{x}{S_+}\right)^2 + C \right]^2
Q_+Q_-}, \nonumber \label{eq:spectrumT}
\end{equation}
with constants
\begin{eqnarray}
 2 H &=& k (p_{+}^2+p_{-}^2) +
\sqrt{k^2 (p_{+}^2 - p_{-}^2)^2 
  +4 (k-1)^2 \left( p_{+}  p_{-} \right)^{2} }, \nonumber \\
 C  &=& k^2 (k-1)^{-1} H^{-1} \left[   
(p_{+}^2+p_{-}^2)H - 2 \left( p_{+}  p_{-} \right)^{2} \right],  \nonumber \\
W&=&\left[H^2 + (2k-1)(p_{+}p_{-})^2 \right], \nonumber \\
Q_{\pm} &=& H \pm (2k-1)p_{+}p_{-}, \nonumber \\
  S_{\pm}^{2} &=& Q^{2}_{\pm} \left[
(H \mp p_{+}p_{-} )^2 - H C \right ]^{-1}.\nonumber 
\end{eqnarray}

\section{\sectionsize Stability analysis: derivation of the support}
The support is found through a stability analysis around the trivial solution
(\ref{eq:3}-\ref{eq:4}).
We make therefore an expansion around the trivial solution of the kind
\begin{eqnarray}
 a &=& a_1, \\ 
b &=& b_1, \\ 
c &=&c_0 + c_1, \\ 
d&=&d_0 + d_1,
\end{eqnarray}
with $|a_1|\ll 1$, $|b_1|\ll1$, $|c_1|\ll1$ and $|d_1|\ll1$.    Substitution in
equations (\ref{eq:GiT}-\ref{eq:GiCT}), this linear
stability analysis leads to the following equation: 
\begin{eqnarray}
 \left|(c^2_0+d^2_0)\right| = H^{-1} \label{eq:supportT}
\end{eqnarray}
The expression in (\ref{eq:support}) is in fact an ellipse of the kind: 
\begin{eqnarray}
x^2 \: Q^{-2}_+ +
y^2 \: Q^{-2}_{-} < H^{-1},
\label{supportTT}
\end{eqnarray}
for $\theta =0$, while for $\theta>0$ the ellipse rotates around the origin by
an angle $\theta$.  This is the formula presented in our main paper.  One can
see that Eq.~(\ref{eq:supportT}) represents an ellipse rather quickly by
considering
that the conformal map 
$\zeta = \lambda \pm \sqrt{\lambda^2-d}$ maps a circle in the
$\zeta$-space on an ellipse in the $\lambda$-space.  Considering the formula
for $c_0$ and $d_0$, Eqs.~(\ref{eq:c0}-\ref{eq:d0}), we notice indeed that
(\ref{eq:supportT}) represents an ellipse.

}

\end{document}